\documentclass{physeauth}

\usepackage[dvipdf]{graphicx}% Include figure files
\usepackage{dcolumn}% Align table columns on decimal point
\usepackage{bm}% bold math
\usepackage{amssymb}

\begin{document}
\begin{frontmatter}
\title{Spectroscopy of soft modes and quantum phase transitions in coupled electron
bilayers}
\author[ad3,ad1]{S. Luin\corauthref{SL},}
\author[ad0,ad1]{I. Dujovne,}
\author[ad3]{V. Pellegrini,}
\author[ad0,ad1,ad2]{A. Pinczuk,}
\author[ad1]{B.S. Dennis,}
\author[ad4]{A.S. Plaut,}
\author[ad1]{L.N. Pfeiffer,}
\author[ad1]{K.W. West,}
\author[ad3]{J.H. Xu}
\address[ad3]{NEST-INFM and Scuola Normale Superiore, Pisa I-56126 Italy}
\address[ad0]{Dept. of Applied Physics and Applied Mathematics , Columbia
 University, New York, NY 10027, USA}
\address[ad1]{Bell laboratories, Lucent Technologies, Murray Hill, NJ 07974, USA}
\address[ad2]{Dept. of Physics, Columbia University, New York,  NY 10027, USA}
\address[ad4]{Dept. of Physics, Exeter University, Exeter, UK}
\corauth[SL]{email: s.luin@nest.sns.it; phone ++39-050 509429; fax ++39-050 509417}
\begin{abstract}
Strongly-correlated two-dimensional electrons in coupled
semiconductor bilayers display remarkable broken symmetry
many-body states under accessible and controllable experimental
conditions. In the cases of continuous quantum phase transitions
soft collective modes drive the transformations that link distinct
ground states of the electron double layers. In this paper we
consider results showing that resonant inelastic light scattering
methods detect soft collective modes of the double layers and
probe their evolution with temperature and magnetic field. The
light scattering experiments offer venues of research of
fundamental interactions and continuous quantum phase transitions
in low-dimensional electron liquids.
\end{abstract}
\end{frontmatter}
\thispagestyle{plain}
\maketitle
\section{Introduction}
The energy $\Delta$ of the lowest excitation above the quantum
ground state represents a characteristic scale for the quantum
transformations in the ground states of strongly-correlated
low-dimensional electron systems at {\it zero temperature}. In
continuous quantum phase transitions (QPTs), $\Delta$ tends to
zero when the critical point of the phase transformation is
approached. The collapse of $\Delta$ at the critical point is
associated with a diverging characteristic length scale giving
rise to specific scaling laws and universal exponents that are
explored experimentally. Significant examples are the scaling
behavior of the resistivity in quantum Hall regimes
\cite{sondhi97}, and of the conductance in semiconductor quantum
dots in the Kondo regime \cite{kastner}.
\par
Several material systems exhibit continuous QPT that are under
current intense investigation \cite{sachdev}. The two-dimensional
electron systems in semiconductor quantum structures display
complex phase diagrams linked to QPTs. Quantization of in-plane
kinetic energy into massively degenerate Landau levels in the
quantum Hall regimes create ideal environments where QPTs can be
induced and studied under controlled experimental conditions. The
coupled electron bilayers realized in wide quantum wells or in
coupled double quantum wells (DQW) display diverse groups of
broken-symmetry states that are linked by QPTs \cite{mac97}. The
richness of the phase diagram of bilayer systems, due to the
interplay between intra- and inter-layer Coulomb interactions, has
led to the discovery of some of the most intriguing behaviors in
contemporary condensed-matter physics \cite{girvinPT}.
\par
In electron double layers different QPTs can be induced at zero
Kelvin by changing the ratio between bare Zeeman energies $E_{z}$
and tunnelling gaps $\Delta_{SAS}$,  where the $\Delta_{SAS}$ are
the spacings between the lowest symmetric and antisymmetric levels
of the electron bilayers. Additional tuning of the bilayer phase
diagram is achieved by changes in the total electron density, the
distance $d$ between the layers and the in-plane component of the
magnetic field \cite{mac97,vitssc01,tam,Murph94,zheng97}. The
relevant low-lying (or soft mode) excitations are here charge- or
spin-collective modes associated with transitions across the
tunnelling gap of the double-layer system. The mode energies have
characteristic dispersions that are determined by the electron
interactions. Excitonic Coulomb interactions (or vertex-correction
terms) of the neutral quasiparticle-quasihole pairs in the
transitions are responsible for minima in the dispersion
that occur at zero or finite in-plane wavevectors $\bf q$.
\par
Inelastic light scattering results presented here highlight
collective mode softening due to excitonic Coulomb interactions.
The mode softening may drive instabilities in the electron
bilayers. Direct access to soft excitations by inelastic
scattering methods offers powerful venues for probing the physics
of QPTs in strongly-correlated systems.  We discuss here results
of inelastic light scattering experiments that probe soft
collective modes in the bilayer electron gas. The experiments
uncover direct links between mode softening and excitonic
interactions. These soft collective modes, while observed in the
symmetric phases, seem to be associated with the emergence of
broken-symmetry states in the electron bilayers. These
experiments, however, are very challenging. We recall that the
observation of soft collective excitations at the superfluid
transition in $^4$He by inelastic neutron scattering
\cite{svensson} is one of the few examples reported in the
literature.
\par
The modulation-doped DQWs studied in our experiments were grown by
molecular beam epitaxy. They consist of two nominally identical
GaAs quantum wells of width $d_w$=18 nm separated either by
~Al$_{0.2}$Ga$_{0.8}$As or ~Al$_{0.1}$Ga$_{0.9}$As barriers with
width ranging from $d_B$ = 4 nm to $d_B$ = 8 nm. Si $\delta$-doped
layers grown both above and below the DQW create two equivalent
two-dimensional electron gases (2DEGs). The combined total carrier
concentrations used in the experiments reported in this paper are
in the range 0.15 $\times$ 10$^{11}$cm$^{-2}<$~n~$<$1.7 $\times$
10$^{11}$cm$^{-2}$ and low-temperature mobilities 0.2 -- 1.5
$\times$ 10$^6$cm$^2$/V$s$. Resonant inelastic light scattering
spectra were obtained in a conventional backscattering geometry at
temperatures between 1.7K and 50 mK using a dye laser tuned to the
optical transitions of  the GaAs quantum wells. Discussions on
fundamentals of resonant inelastic light scattering processes by collective excitations,
matrix elements, resonant enhancements and conservation laws can
be found in Refs. \cite{vitssc01,pin97,hawr85,tsel84}. Magnetic fields $B_{T}$
are applied at angles $\theta$ with the normal to the bilayers.
The perpendicular component of the field is $B = B_{T}\cos\theta$.
\par
In Section 2 we review the results at $B$=0 that set up a
conceptual and practical framework for studies of unstable $\bf q$
= 0 spin excitation modes at filling factor $\nu = 2$.  In Section
3 we consider recent results obtained in the vicinity of $\nu =2$
at temperatures below 100mK that support the notion that a $\bf q$
= 0 spin soft-mode drives a QPT in electron bilayers. In Section 4
we discuss the most recent experiments at total filling factor
$\nu =1$. These results show that the light scattering experiments
provide evidence of soft tunnelling charge-density excitations at
finite $\bf q$ (soft magneto-rotons) in close proximity to the 
compressible-incompressible phase boundary.
\par
\section{Coupled bilayers at B=0}
Soft-mode driven QPTs were predicted to occur at $B$=0 \cite{tam}.
In these theories, the instability is caused by a
vertex-correction driven softening of the {\bf q}=0 spin-density
tunnelling excitation (SDE) across the two lowest subbands
represented by the $\Delta_{SAS}$ gap. At zero magnetic field this
excitation is a degenerate triplet characterized by changes of
$\delta S_z$ = $\pm 1$ and $\delta S_z$ = 0 in spin angular momentum
along the normal to the bilayers . The impact of the vertex
correction is readily seen in the expression for the mode energy
\cite{dec}
\begin{equation}
\omega_{SDE}^2 = \Delta_{SAS}^2 - 2 (n_S - n_{AS}) \, \beta \, \Delta_{SAS},
\label{eq:wSDE}
\end{equation}
where $n_S$  and $n_{AS}$ are the populations of the symmetric and
antisymmetric subbands.  The factor $(n_S-n_{AS})$ takes into
account the reduction in  available phase space when two subbands
are occupied. The second  term in the right-side of Eq.\
(\ref{eq:wSDE}) represents the excitonic binding  and the factor
$\beta$ parametrizes the strength of vertex corrections. When this
intrawell excitonic term is larger than the interwell couplings
represented by $\Delta_{SAS}$,  the {\bf q}=0 SDE collapses. The
possible existence of such unstable spin-density excitations
suggests the emergence of broken-symmetry phases of
antiferromagnetically correlated 2D layers. The time-dependent
local-density approximation (TDLDA) results of Ref. \cite{tam}
predict a phase transition for electron densities below
0.5$\times$10$^{11}$cm$^{-2}$.
\par
In this section we present the results of a comprehensive search
for the soft {\bf q}=0 intersubband SDEs at zero magnetic field.
While this work demonstrates the absence of such instabilities,
it provides the framework for the design and interpretation of
the experiments at non-zero magnetic fields presented in the next
section. Resonant  inelastic light-scattering measurements of the
SDE modes have been  carried out on a large set of GaAs DQWs
of various well shapes, and with electron
densities down to 10$^{10}$cm$^{-2}$. The  sample parameters were
chosen to overlap those for which TDLDA calculations indicate the
existence of an unstable  SDE that triggers transitions to
broken-symmetry phases.
\par
Figure~1 shows typical lowest energy intersubband spectra. For
{\bf q}$\simeq$0, the energy of the single-particle excitations
(SPE) equals that of $\Delta_{SAS}$.  The highest energy peak
arises from the CDE. The CDE energy is given by \cite{dec}
\begin{equation}
\omega_{CDE}^2 = \omega_{SDE}^2 + 2 (n_S - n_{AS}) \, \alpha \, \Delta_{SAS}
\label{eq:wCDE}
\end{equation}
where $\alpha$ is the depolarization shift due to direct terms of
Coulomb interactions. Usually $\alpha > \beta$, so that the CDE
appears higher in energy than both the SPE and SDE \cite{dec}. The
collective SDE and CDE display well-defined polarization selection
rules.  SDE are active in depolarized spectra, where incident and
scattered light polarizations are perpendicular and CDE occur in
polarized spectra measured with parallel polarizations.

\begin{figure}[ht]
\begin{center}
\includegraphics[width=0.8\columnwidth]{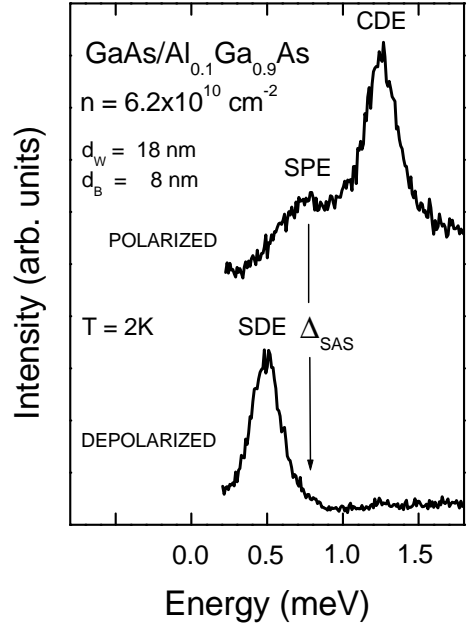}%
\label{annette1}
\caption{
\baselineskip 24pt
\noindent
Inelastic light-scattering spectra of the intersubband excitations of the
double quantum well. The peaks of the spin-density excitations (SDE),
charge-density excitations (CDE) and single-particle excitations (SPE)
are shown. The spectra have been offset for clarity.
}
\end{center}
\end{figure}

We have measured the energy of the CDE, SDE and SPE modes in a series of
samples as a function of carrier concentration. Results are plotted in Fig.2(a).
A striking feature of the TDLDA calculation  is the vanishing SDE mode energy
for electron densities below 0.5$\times$10$^{11}$cm$^{-2}$ \cite{tam}.
This prediction of the LDA is absent in the
experimental results  of Fig.~2a.
\begin{figure}[bt]
\begin{center}
\includegraphics[width=0.8\columnwidth]{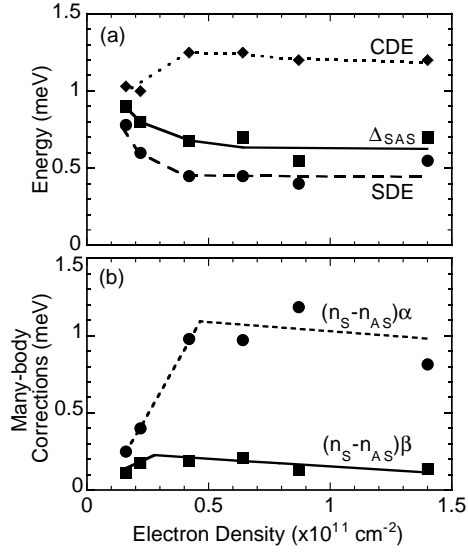}%
\label{annette2}
\caption{
\baselineskip 24pt
\noindent
The measured two-dimensional electron density dependence of (a)
the intersubband charge-density excitation energy $\omega _{CDE}$
(dotted), spin-density excitation energy $\omega _{SDE}$ (dashed) and
single-particle excitation energy $\Delta_{SAS}$ (solid); and
(b) $(n_S~-~n_{AS}) \alpha$ (dashed), and
$(n_S~-~n_{AS}) \beta$ (solid), for the
GaAs/Al$_{0.1}$Ga$_{0.9}$As double quantum well structures.
The lines are guides to the eye.
}
\end{center}
\end{figure}

Other features predicted by LDA calculations are observed
experimentally. We find good agreement between the experimental
and the LDA values of SPE and CDE energies.  The experimental SPE
energies show the upward trend with decreasing electron density
expected from the  renormalized $\Delta_{SAS}$  calculated self-
consistently within LDA \cite{mss}. In addition, LDA calculations
predict that at n$<$10$^{10}$cm$^{-2}$ the excitonic vertex
correction should become larger than the depolarization shift
resulting in the energy of the CDE falling below that of the SPE
\cite{marm,goni}.  And indeed, in the experimental data a  similar
decrease in the CDE energy is observed as n approaches
10$^{10}$cm$^{-2}$. As will be seen below, this is due to a
decrease in the depolarization shift rather than an increase in
the excitonic shift, which is again in  qualitative agreement with
LDA theory \cite{marm}.
\par
To obtain further insight into the many-body interactions we can rewrite Eqs.\
(\ref{eq:wSDE}) and  (\ref{eq:wCDE}) to determine the direct and exchange terms
of the Coulomb interaction: The depolarization shift, $(n_S-n_{AS}) \alpha$, and
the excitonic shift, $(n_S-n_{AS}) \beta$,  derived from the measured mode
energies shown in Fig.~2a,  are plotted in Figure~2b.  LDA calculations
\cite{briefreport} predict that  $(n_S~-~n_{AS}) \alpha$  increases slightly with
decreasing n and then drops sharply for electron densities  below
0.2~x~10$^{11}$cm$^{-2}$. The maximum  corresponds to the density at which the
second subband just becomes occupied. In Figure~2b, we do indeed observe this
behaviour experimentally. Thus the LDA seems to work well when describing the
depolarization shift down to very low electron densities.
\par
A different situation prevails in the case of the
vertex-correction  $(n_S~-~n_{AS}) \beta$ because here we  find
significant quantitative differences between experiment and LDA
theory:  For electron densities below 0.5~x~10$^{11}$cm$^{- 2}$
the experimental values of $(n_S~-~n_{AS}) \beta$ are 2~$-$5 times
smaller than the LDA would predict, with the most significant
discrepancies occurring at densities below
0.2~x~10$^{11}$cm$^{-2}$,  close to the onset of occupation of the
antisymmetric state.
\par
Thus, TDLDA  is not successful at describing  the observed behaviour of both the
vertex-corrections and the SDE  mode energies at very low density.  We attribute
these discrepancies, between experiment and LDA theory,  to shortcomings of the
TDLDA calculation at low densities, when  local-density approximations  become
unreliable because  very dilute electron gases can no longer be considered
homogeneous even on a local scale.  On the other hand, a non-local approach,
that treats the exchange  Coulomb interaction by means of a variational solution of the
Bethe-Salpeter equation in the ladder approximation \cite{gam,briefreport}, yields
predictions for the energies of  intersubband collective modes in the DQW that are in
good  agreement with experiment.
\par
The absence of zero-field enhancements of  the vertex-correction $(n_S~-
~n_{AS})\beta$, determined experimentally, indicates that at zero field, and at small
values  of $\Delta_{SAS}$, the effective electron density, given by
$(n_S~-~n_{AS})$, is limited by the occupation of the  antisymmetric state.
However, these results are intriguing because, as described below,  pronounced
softenings of low-energy {\bf q}=0 SDE, in general agreement  with Eq.\
(\ref{eq:wSDE}), are  seen in spectra from DQWs in perpendicular  magnetic fields at
even values of $\nu$. Our understanding of this is that the softenings of the
spin-density mode in perpendicular field, described below, are due  to changes in the
effective density of electrons that contribute to the intersubband collective excitations.
This effective density can be changed by magnetic field, and for filling factor $\nu$=2
and lower all the electrons can and do contribute.

\section{Coupled bilayers at $\nu =2$}
\begin{figure}
\begin{center}
\includegraphics[width=0.85\columnwidth]{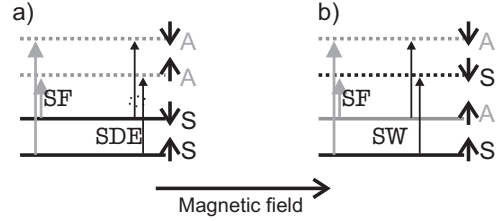}
\caption{\label{levels} Schematic representation of the energy level scheme at
$\nu$=2. a) In the unpolarized state (phase U), when the symmetric states are populated and
b) spin polarized state (phase P), with the lowest symmetric and antisymmetric levels fully occupied.
The arrows correspond to the possible excitations in these configurations. SF: spin flip mode; SDE: spin-density excitation; SW: spin wave. The arrows on the side of the levels correspond to spin.}
\end{center}
\end{figure}

At this even-integer value of Landau level filling factor, the
spin configuration depends on the relationship between $E_{z}=
g\mu_{B} B_{T}$ and $\Delta_{SAS}$ ($g\approx -0.44$ is the gyromagnetic 
factor and $\mu _{B}$ is the Bohr magneton), in conjunction with changes in
available phase space for the tunnelling intersubband transitions
at $\nu =2$. There are two possible configurations of spin (see
Fig. \ref{levels}). A spin-singlet quantum Hall paramagnet (phase
U) is obtained when the electrons occupy the lowest spin-split
symmetric levels. In this configuration the spin polarization is
zero and the lowest-energy spin transitions are spin-flip (SF) and
SDE modes across $\Delta_{SAS}$. Contrary to the $B$=0 case, these
excitations are not degenerate under the application of a magnetic
field. In the Hartree-Fock framework their energies are $\omega
_{SDE}$ for SDE modes and $\omega_{\pm}=\omega_{SDE}-E_z\delta
S_z$, $\delta S_{z} = \pm 1$ for spin-flip (SF) excitations.
%\par
The other configuration possible is a spin-polarized quantum Hall
ferromagnet (phase P). Characteristic excitations in this
configuration are spin waves (SW) across the Zeeman gap and SF
modes.
\par
The actual spin configuration depends on the strength of the
energy required to produce a spin-flip relative to that of the
interlayer interactions that enter in the $\Delta _{SAS}$. The
energy required to change the orientation of spin incorporates two
terms. One is the Zeeman energy $E_{z}$. The other is the spin
stiffness of the 2DEG, given by the increase in Coulomb exchange
interaction energy when the spin orientation is changed. Because
the spin stiffness has a characteristic $B^{1/2}$-dependence, the
spin-singlet configuration prevails at the lower values of $B_T$.
The spin-polarized configuration emerges at higher total fields
when the spin-flip transition energy in the lowest Landau level is
larger than the energy associated with a transition across the
tunnelling gap.
\par
Contrary to predictions and experimental observations of
first-order phase transitions between phases U and P in single
layers \cite{quinn1,wu85,Piazza}, inter-layer interactions in the
bilayer systems leads to a novel broken-symmetry intermediate
phase D linked to phase U by a continuous second-order phase
transition \cite{vitssc01,zheng97,vitsci,quinn2}. Under the
application of parallel magnetic fields, the ratio between the
tunneling and Zeeman gaps can be changed and the system can make
transitions from the singlet phase U to the intermediate
phase D and at larger parallel fields, to phase P. Signatures of
these different configurations of spins were obtained by inelastic
light scattering measurements of spin tunneling excitations
\cite{vitssc01,vitsci}. These works demonstrated that the
transition between U and D is associated to the softening of the
tunneling SDE. The softening is caused by excitonic interactions
that reduce $\omega_{SDE}$ energies to values much lower than
$\Delta_{SAS}$ and close to $E_z$ signaling the existence of an
unstable SF excitation with $\delta S_z$ = 1. In addition to
removing the degeneracy of the three spin excitations in the
triplet state, the application of the magnetic field allows to
manipulate the phase-space factor $(n_S-n_{AS})$. At total filling
factor $\nu=2$ when all the electrons occupy the symmetric states,
we have $n_S-n_{AS} =$n (where n is the total electron density)
and all electrons can contribute to the tunnelling transitions
\cite{vitprl}.
\par
The zero-temperature properties of the ground state in the new
intermediate phase have been the subject of intense theoretical
investigations
\cite{zheng97,dassarma97,demler99,macdonald99,yang99,demler00}. It
is now well accepted that the competition between interlayer
tunnelling, Zeeman splitting, intralayer and interlayer Coulomb
interactions, leads to a spontaneous symmetry breaking in the
bilayer and electron spins in each layer tilted away from the
external magnetic field direction (canted phase).
\par
We focus here on the SDE behavior observed under changes of $B_T$
while keeping the filling factor of $\nu=2$. Detailed analysis of
the SDE softening down to T=200mK when the $\nu$ = 2 state is
approached in a tilted-field configuration indicates that the SDE
energy reaches its lower values in the low-field side of $\nu=2$
(and then disappears for larger values of magnetic field) but at
energies significantly larger than $E_z$ (implying a spin-flip
mode at energy $\omega _{-}$ above zero) \cite{vitssc01,vitsci}.
In order to fully understand these results, it is important to
recall that in the T$>$0 region close to the critical point of
QPTs, large thermal population of low energy soft modes cause
critical fluctuations that destroy the long-range order of the
broken symmetry phase \cite{sondhi97}. In the case of the
continuous QPT in coupled bilayers at $\nu $=2 this could imply
that at T$>$0, the system become unstable before reaching $\nu$=2
and $\omega_{SDE}$ = $E_z$. This is consistent with results shown
in Refs.\cite{vitssc01,vitsci}. In a attempt to further validate
this picture we have extended the inelastic light scattering
measurements at lower temperatures (down to 50mK). At the lowest
temperature, the energies are smaller than at 300mK. The fact that
the energy approaches $E_z$ when the temperature is lowered
provides evidence that the transition is almost continuous.
\par
\begin{figure}
\begin{center}
\includegraphics[width=0.75\columnwidth]{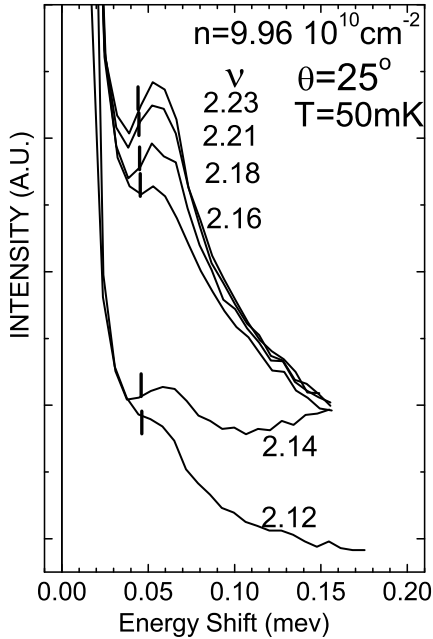}
\caption{\label{fig-spectra}
Typical spectra at different filling factors for $\nu >2$. Experiments were done at
50mK and a tilt angle of 25$^o$. The vertical lines on each spectra correspond to the value of the
Zeeman energy.}
\end{center}
\end{figure}

 Figure \ref{fig-spectra} shows typical spectra at different filling factors near
$\nu$=2 and T=50mK. The sample (a DQW with 8 nm
~Al$_{0.1}$Ga$_{0.9}$As barrier, density of $9.96\times 10^{10}
$cm$^{-2}$ and mobility close to $10^6 $cm$^{2}/$V$s$) is tilted
(25$^o$) from the plane perpendicular to the magnetic field. There
are two main contributions to these spectra: the sharp peak
corresponds to the inelastic light scattering signal due to the
SDE, while the broad background is due to
photoluminescence.
\par
\par
\begin{figure}
\begin{center}
\includegraphics[width=0.86\columnwidth]{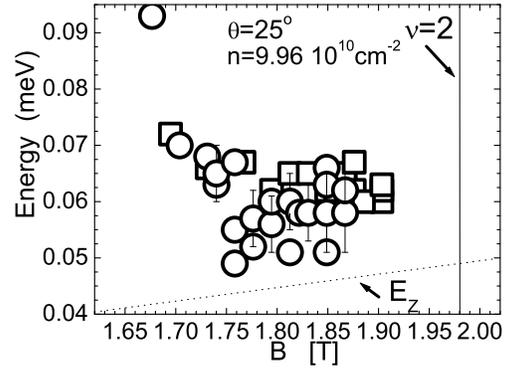}
\caption{\label{fig-25} Magnetic field dependence of the spin-density excitation
in the vicinity of $\nu=2$ at 50mK ($\bigcirc$) and 300mK ($\Box$).
The experiments were done at a tilt angle of 25$^o$. Typical error bars are shown.}
\end{center}
\end{figure}

The impact of temperature changes near $\nu=2$ is shown in Fig.
\ref{fig-25}. Previous results \cite{vitssc01,vitsci} have
demonstrated that at 300mK and at tilt angle of 20$^o$ the system
is in the spin unpolarized phase, while at 35$^o$ the sample is
already in the ferromagnetic phase. SDE energies are displayed at
different magnetic fields near $\nu=2$ for 50mK (circles) and
300mK (squares) and at 25$^o$. At our lowest temperature
attainable, 50mK, the system shows the same type of behavior of
the SDE peaks seen at 300mK with same angle, but consistently at
lower energies. The excitation softens markedly as the filling
factor is diminished (larger magnetic fields), and disappears as
it gets closer to $\nu$=2 (vertical line in Fig. \ref{fig-25}). In
these regions of magnetic fields the energy of the mode is very
close to $E_z$ (see dotted line in Fig. \ref{fig-25}). It is
possible that the mode becomes overdamped very close to $\nu$=2,
and this could explain why the mode is no longer observable at
filling factors slightly larger than $\nu$=2, where its energy is
comparable to $E_z$. Detailed values of the Zeeman energy $E_z$
are obtained from inelastic light scattering measurements of SW
excitations in phase P.
\par
The softening of the intersubband SDE to values close to $E_z$
implies the emergence of spin flip excitations with $\delta S_z$=1
at vanishing small energies that could lead to instabilities in
the system. The temperature dependence (down to the lowest
temperatures we can achieve) shows that the modes behave in a
manner consistent with a QPT at T=0.
\par
\section{Coupled bilayers at $\nu =1$}
In the previous section we have shown that $\bf q$ = 0 spin
tunnelling modes in coupled bilayers can become unstable. Unstable
soft modes at finite wavevectors were also predicted to occur in electron-hole
systems at strong magnetic fields \cite{chen91} and in electron bilayers at
total filling factor $\nu $ = 1 \cite{Fert89,MacD90,Brey90,chen92,Jogl02}.
In this section we consider resonant inelastic light scattering studies
that probe such excitations.
\par
We recall that the relevant range of in-plane wavevectors ($\bf
q$) accessible to inelastic light scattering experiments is
restricted by wavevector conservation in the scattering
process to scattering wavevectors $q$ $\approx $10$^{5}$ cm$^{-1}$.
This wavevector range implies access to mode wavevectors $q\le
0.2l^{-1}_B$, where $l_B$ is the magnetic length. For this reason
the measurements discussed in the previous sections access the
long-wavelength limit ${\bf q}\rightarrow 0$. The conservation law
of wavevector originates from the translational invariance of the
2D system. Even the most perfect samples of 2D electron systems
are subject to residual disorder, particularly in the quantum Hall
regimes. One of the significant manifestations of residual
disorder in resonant inelastic light scattering is the activation
of large wavevector collective modes \cite{pinczuk88,Marm92}. The
experiments at $\nu = 1$ take advantage of this feature of light
scattering experiments to study soft magnetoroton modes.
\par
A distinct feature of the Hartree- Fock calculations is the
magnetoroton minimum at finite wavevectors ${q}\sim l^{-1}_{B}
$\cite{Kall84}. Magnetoroton excitations were predicted to occur
in the dispersions of inter-Landau level modes and in intra-Landau
level excitations of the fractional quantum Hall regime
\cite{Kall84,girvin85,girvin86}. The minima arise, as mentioned
above, from the $q$-dependent excitonic binding in
quasiparticle-quasihole pairs.
\par
Observation of roton minima by resonant inelastic light scattering
requires breakdown of wavevector conservation assigned to the
impact of residual disorder. With absence of wavevector
conservation, the light spectra can reveal features of the density
of states of Landau level excitations. Massive breakdown of
wavevector conservation was indeed invoked to interpret resonant
inelastic light scattering experiments in the integer and
fractional regimes \cite{pinczuk88,davies97,Kang01}. These
experiments revealed magnetoroton excitations and excitations with
$q\gg l^{-1}_B$.
\par
We have taken advantage of breakdown of wavevector conservation in
resonant inelastic light scattering to explore possible links
between soft magnetorotons and QPTs in electron bilayer in the
quantum Hall regimes \cite{Luin03}. Soft rotons are crucial
elements in the theoretical constructions that predict and
interpret QPTs in the two-dimensional electron system. One
remarkable example is given by the transition to a Wigner crystal
state predicted to occur at fractional filling factors beyond 1/7
\cite{girvin85,jain}. This link, however, has not yet been
addressed in experiments.
\par
\begin{figure}[bt]
\begin{center}
\includegraphics[width=0.87\columnwidth]{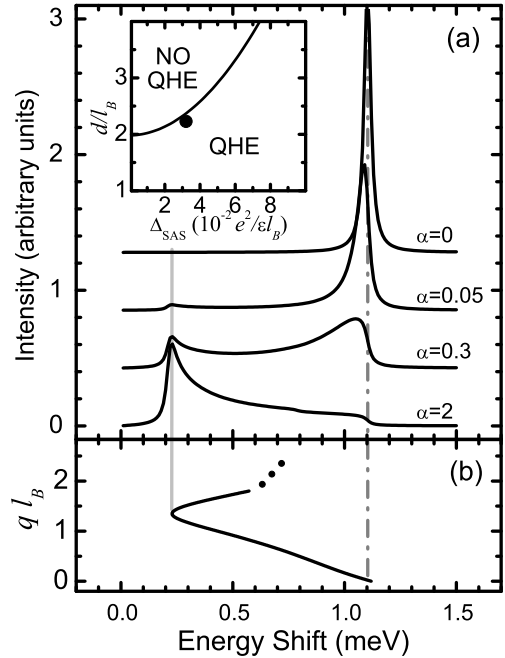}%
\caption{\label{luin1} Panel (a): Calculated inelastic light
scattering spectra. $\alpha $ labels the breakdown of wavevector
conservation (see text). The curves are the response functions
calculated in TDHFA including the wavevector conservation
breakdown, using the dispersion shown in panel (b), for a sample
close to the phase transition (phase diagram in inset of panel
(a))}
\end{center}
\end{figure}

The role of soft magnetorotons is particularly intriguing in
coupled bilayers at total Landau level filling factor $\nu_T =1$.
At this filling factor value, the bilayer system displays a rich
phase diagram determined by the interplay between single-particle
tunnelling gap $\Delta_{SAS}$ with intra- and inter-layer
interactions\cite{mac97}. Two distinct phases were identified on
the basis of the value of the longitudinal magneto-resistance
measured in magneto-transport experiments. One of the two phases
is characterized by the existence of the quantum Hall effect at
$\nu_T =1$ while the other not. The phase diagram is shown in the
inset of Fig.~\ref{luin1} \cite{Murph94,Boebinger}. In current
theories the compressible-incompressible QPT is linked to a roton
instability in the CDE across the tunnelling gap
\cite{Fert89,MacD90,Brey90,Jogl02}. In Hartree-Fock calculations
the roton softening instability is due to large excitonic bindings
between quasiparticle and quasihole that occurs with a reduction
in $\Delta_{SAS}$.
In order to show how the breakdown of wavevector conservation can
lead to manifestations of collective CDE modes at finite $q$ in
the inelastic light scattering spectra we first recall that
intensities of inelastic light scattering is in the lowest order
approximation proportional to the dynamic structure factor
$S\left(q,\omega;\alpha\right)$. Here $\alpha$ is a
phenomenological broadening parameter in wavevector space, used to
account for the effects of disorder in the breakdown of wavevector
conservation. Within this model, used by Marmorkos and Das Sarma
\cite{Marm92},  we have:
\begin{equation} \label{eqSS}
S\left(q,\omega;\alpha\right) \sim
\frac{\alpha}{\pi l_B}\int\! dq'
\frac{S\left(q',\omega\right)}{\left(q-q'\right)^2+\left(\alpha\, l_B^{-1}\right)^2}\,,
\end{equation}%
where $S\left(q',\omega\right)$ is the electronic dynamic
structure factor in the translation invariant system.
$S\left(q',\omega\right)$ for charge-density excitations across
the symmetric-antisymmetric gap is given by:
\begin{equation} \label{eqS}
S\left(q,\omega\right) \propto
\frac{\left|M\left(q\right)\right|^2\omega_C\!\left(q\right)\,\omega\Gamma}
{\left[\omega^2-\omega_C^2\!\left(q\right)\right]^2+\omega^2\Gamma^2}\,,
\end{equation}%
where $\Gamma$ is a homogeneous broadening.
\par
To highlight the impact of breakdown of wavevector conservation we
focus on a sample in close proximity to the phase boundary (see
the dot in the inset of Fig.~\ref{luin1}(a)) for which the
calculated dispersion is shown in Fig.~\ref{luin1}(b). The
dispersion is characterized by a deep magnetoroton which
anticipates the occurrence of a continuous QPT. We note that TDHFA
being a mean-field theory fails to reproduce the collective mode
dispersion when $q\gg l^{-1}_B$. That part of the dispersion was
not included in the model. Figure~\ref{luin1}(a) shows the
calculated inelastic light scattering spectra for different values
of $\alpha$.  In these calculations a broadening factor
$\Gamma=0.035\,$meV, due to lifetime limited by disorder and
impurity scattering, was considered \cite{Marm92}. Further details
on the results of the calculations can be found in
Ref.\cite{Luin03}. In Fig. ~\ref{luin1}(a) we find that small
values of $\alpha$, of the order of $0.1 l_{o}^{-1}$ , are
sufficient to yield significant intensity at the energy of
magnetoroton critical point in the density of states.  This is due
to the fact that the matrix element $|M(q)|^2$ that enters the
dynamic structure factor and acts as the oscillator strength for
inelastic light scattering tends to peak sharply at the
magnetoroton wavevector when the bilayer approaches the phase
transition instability \cite{Luin03}.
\begin{figure}
\begin{center}
\includegraphics[width=0.9\columnwidth]{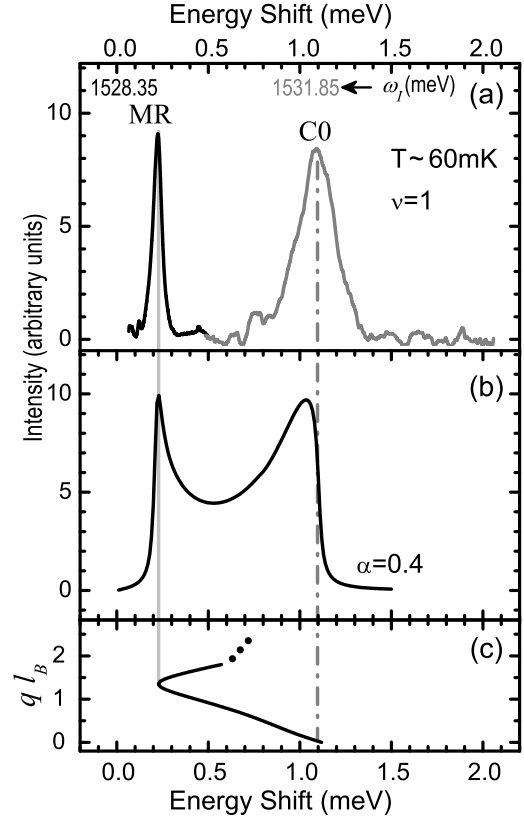}%
\caption{\label{luin2} Comparison within theory and experiment. Panel (a): MR and C0 resonant Raman peaks, for two different incident photon energies (reported in the figure). Background is subtracted. Panel (b): response function for a chosen value of $\alpha $=0.4. (c): collective mode dispersion.}
\end{center}
\end{figure}%
\par
In Fig. ~\ref{luin2} we compare measured spectra
(Fig.~\ref{luin2}(a)) with calculations in which $\alpha$ is set at 0.4
(Fig.~\ref{luin2}(b)). The sample is characterized by a
7.5-nm-wide $Al_{0.1}Ga_{0.9}As$ barrier, electron density
n=$1.2\times 10^{11}cm^{-2}$ and $\Delta_{SAS} = 0.32$ meV and its
position in the phase diagram corresponds to the dot in the inset
of Fig.~\ref{luin1}. A value of $\alpha = 0.4$ yields an effective
length scale of $\sim 3l_B$ for the terms in the disorder potential
responsible for the breakdown of wavevector conservation that
interprets the data. The resonant inelastic light scattering
spectra are obtained at two different incident photon energies and
displayed after conventional subtraction of background due to
laser and magneto-luminescence. The assignments of the two
spectral features to CDE at $q=0$ (C0) and to the magneto-roton
(MR) are consistent with the calculated dispersion shown in
Fig.~\ref{luin1}(b) and Fig.~\ref{luin2}(c).
\par
The calculation reproduces the energy position and the widths of
the peaks. The fact that the C0 width is larger than that of the
MR is significant and seems to reflect the peculiar behavior of
the $|M(q)|^2$ factor close to the phase boundary. On the
contrary, results obtained from a sample far from the phase
boundary show a broad and weak MR peak (compared to C0) and
located at higher energies ($\approx 0.6$ meV) \cite{Luin03}. The
calculated spectra shown in Fig.~\ref{luin2} differ from the
measurements in the region between the C0 and MR modes. The
discrepancy could be related to differences in resonant
enhancements.

\section{Conclusions and Perspectives}
We have presented extensive results that use light scattering
methods to probe soft, unstable, collective excitations in 
two-dimensional electron systems. Inelastic light scattering 
has the capability to detect collective soft modes at
finite wavevectors as well as spin- or charge-density modes at
$\bf q$ = 0  and at $\bf{q} = \infty$. Since the latter are the
only excitations probed in magneto-transport, light scattering
studies have the potential to bring a new wealth of information
into studies of quantum phase transitions, spin states and other
new broken-symmetry phases in modulation-doped low-dimensional
semiconductors. The observations at zero magnetic fields and at
total filling factor $\nu $ = 2 offer significant examples of the
capabilities of this approach. The experiments at $\nu=2$ are the
first to demonstrate mode softening due to excitonic Coulomb
interactions in the quantum Hall regime. The experiments at total
filling factor $\nu $ = 1 are particularly revealing since they
offer direct evidence of soft magnetorotons in the quantum Hall
regime. These results are significant because they suggest that
roton softening plays major roles in the
incompressible-compressible phase transitions of the bilayers at
$\nu =1$, and suggest a leading role of excitonic Coulomb
interactions in transitions between the highly correlated phases.
The results are consistent with a scenario in which the
instability is associated with the condensation of neutral
excitons, a subject of current interest in bilayers and in other
semiconductor systems \cite{JPE,Butov,Snoke}. The possible link
between exciton condensation and quantum Hall phases in coupled 
bilayers highlights the role of the quantum Hall regime 
as a versatile laboratory, almost ``magic'' in its enormous
possibilities, to study quantum liquid phenomena relevant to a
multitude of different areas of condensed-matter physics. The
liquids of quantum Hall states in semiconductor structures, in fact, 
can be created and probed in controlled experiments. To this end, it is
worth noting the similarities between the magnetoroton observed in
the coupled bilayers with the finite-wavevector excitations
observed by inelastic neutron scattering in superfluid $^4$He
\cite{svensson} and in the high-$T_c$ superconductor
$YBa_{2}Cu_{3}O_{7}$ \cite{Fong}. These links highlight the impact
of experimental and theoretical research on quantum phase
transitions and broken-symmetry states in the bilayer electron
gas.

\section{Acknowledgements}
These works were supported by the Nanoscale
Science and Engineering Initiative of the National Science
Foundation under NSF Award Number CHE-0117752, by a research grant
of the W. M. Keck Foundation, by CNR (Consiglio Nazionale delle
Ricerche), by INFM/E (Istituto Nazionale per la Fisica della
Materia, section E), by the Italian Ministry of
University and Research (MIUR) and by the Leverhulme Trust (UK).

\bibliography{SSC18june}

\begin{thebibliography}{10}

\bibitem{sondhi97}
S.~L. Sondhi and et. al.
\newblock {\em Review of Modern Physics}, 69:315, 1997.

\bibitem{kastner}
M.A. Kastner and D.~Goldhaber-Gordon.
\newblock {\em Solid State Comm.}, 119:245, 2001.

\bibitem{sachdev}
S.~Sachdev.
\newblock {\em Quantum Phase Transitions}.
\newblock Cambridge University Press, 1999.

\bibitem{mac97}
S.~M. Girvin and A.~H. MacDonald.
\newblock {\em Perspectives in Quantum Hall Effects, S. Das Sarma and A.
  Pinczuk editors}, Wiley, New york, Chapter 5:161, 1997.

\bibitem{girvinPT}
S.~Girvin.
\newblock {\em Phys. Today}, page~39, June 2000.

\bibitem{vitssc01}
V.~Pellegrini and A.~Pinczuk.
\newblock {\em Solid State Comm.}, 119:301, 2001.

\bibitem{tam}
S.~Das~Sarma and P.I.~Tamborenea, Phys. Rev. Lett. {\bf 73}, 1971 (1994);
  R.J.~Radtke and S.~Das~Sarma, Solid State Commum. {\bf 96} 215 (1995); Solid
  State Commum. {\bf 98} 771 (1996).

\bibitem{Murph94}
S.~Q. Murphy, J.~P. Eisenstein, G.~S. Boebinger, L.~N. Pfeiffer, and K.~W.
  West.
\newblock {\em Phys. Rev. Lett.}, 72:728, 1994.

\bibitem{zheng97}
Lian Zheng, R.~J. Radtke, and S.~Das Sarma.
\newblock {\em Phys. Rev. Lett.}, 78:2453, 1997.

\bibitem{svensson}
E.C. Svensson, W.~Montfrooij, and I.M. de~Schepper.
\newblock {\em Phys. Rev. Lett.}, 77:4398, 1996.

\bibitem{pin97}
A.~Pinczuk.
\newblock {\em Perspectives in Quantum Hall Effects, S. Das Sarma and A.
  Pinczuk editors}, Wiley, New york, Chapter 8:307, 1997.

\bibitem{hawr85}
P.~Hawrylak, J.~Wu, and J.~J. Quinn.
\newblock {\em Phys. Rev. B}, 32:5169, 1985.

\bibitem{tsel84}
A.~C. Tselis and J.~J. Quinn.
\newblock {\em Phys. Rev. B}, 29:3318, 1984.

\bibitem{dec}
R.~Decca, A.~Pinczuk, S.~Das~Sarma, B.S.~Dennis, L.N.~Pfeiffer and K.W.~West,
  Phys. Rev. Lett. {\bf 72}, 1506 (1994).

\bibitem{mss}
A.~S.~Plaut, A.~Pinczuk, B.S.Dennis, J.P.~Eisenstein, L.N.~Pfeiffer and
  K.W.~West, Solid-State Electronics {\bf 40}, 291 (1996).

\bibitem{marm}
S.~Das~Sarma and I.K.~Marmorkos, Phys. Rev. B {\bf 47}, 16343 (1993);
  I.K.~Marmorkos and S.~Das~Sarma, Phys. Rev. B {\bf 48}, 1544 (1993).

\bibitem{goni}
S.~Ernst, A.R.~Goni, K.~Syassen and K.~Eberl, Phys. Rev. Lett. {\bf 72}, 4029
  (1994).

\bibitem{briefreport}
A.S.~Plaut, A.~Pinczuk, P.I.~Tamborenea, B.S.Dennis, L.N.~Pfeiffer and
  K.W.~West, Phys. Rev. B {\bf 55}, 9282 (1997).

\bibitem{gam}
J.C.~Ryan, Phys. Rev. B {\bf 43}, 12406 (1991); D.~Gammon, B.V.~Shanabrook,
  J.C.~Ryan, D.S.Katzer and M.J.~Yang, Phys. Rev. Lett. {\bf 68}, 1884 (1992);
  S.L.~Chuang, M.S.C.~Luo, S.~Schmitt-Rink and A.~Pinczuk, Phys. Rev. B {\bf
  46}, 1897 (1992).

\bibitem{quinn1}
G.F. Giuliani and J.J. Quinn.
\newblock {\em Phys. Rev. B}, 31:6228, 1985.

\bibitem{wu85}
J.~W. Wu, P.~Hawrylak, and J.~J. Quinn.
\newblock {\em Phys. Rev. B}, 31:6592, 1985.

\bibitem{Piazza}
V.~Piazza, V.~Pellegrini, F.~Beltram, W.~Wegscheider, T.~Jungwirth, and A.H.
  MacDonald.
\newblock {\em Nature}, 402:638, 1999.

\bibitem{vitsci}
V.~Pellegrini, A.~Pinczuk, B.S. Dennis, A.S. Plaut, L.N. Pfeiffer, and K.W.
  West.
\newblock {\em Science}, 281:799, 1998.

\bibitem{quinn2}
G.F. Giuliani and J.J. Quinn.
\newblock {\em Surface Science}, 170:316, 1986.

\bibitem{vitprl}
V.~Pellegrini, A.~Pinczuk, B.S. Dennis, A.S. Plaut, L.N. Pfeiffer, and K.W.
  West.
\newblock {\em Phys. Rev. Lett.}, 78:310, 1997.

\bibitem{dassarma97}
S.~{Das Sarma}, Subir Sachdev, and Lian Zheng.
\newblock {\em Phys. Rev. Lett}, 79:917, 1997.

\bibitem{demler99}
E.~Demler and S.~{Das Sarma}.
\newblock {\em Phys. Rev. Lett}, 82:3985, 1999.

\bibitem{macdonald99}
A.~H. Macdonald, R.~Rajaraman, and T.~Jungwirth.
\newblock {\em Phys. Rev. B}, 60:8817, 1999.

\bibitem{yang99}
Min-Fong Yang and Ming-Che Chang.
\newblock {\em Phys. Rev. B}, 60:13985, 1999.

\bibitem{demler00}
E.~Demler, Eugene~H. Kim, and S.~{Das Sarma}.
\newblock {\em Phys. Rev. B}, 61:10567, 2000.

\bibitem{chen91}
X.~M. Chen and J.~J. Quinn.
\newblock {\em Phys. Rev. Lett.}, 67:895 and 2113, 1991.

\bibitem{Fert89}
H.~A. Fertig.
\newblock {\em Phys. Rev. B}, 40:1087, 1989.

\bibitem{MacD90}
A.~H. MacDonald, P.~M. Platzman, and G.~S. Boebinger.
\newblock {\em Phys. Rev. Lett.}, 65:775, 1990.

\bibitem{Brey90}
L.~Brey.
\newblock {\em Phys. Rev. Lett.}, 65:903, 1990.

\bibitem{chen92}
X.~M. Chen and J.~J. Quinn.
\newblock {\em Phys. Rev. B}, 45:11054, 1992.

\bibitem{Jogl02}
Y.~N. Joglekar and A.~H. MacDonald.
\newblock {\em Phys. Rev. B}, 65:235319, 2002.

\bibitem{pinczuk88}
{A. Pinczuk}, J.P. Valladares, D.~Heiman, A.C. Gossard, J.H. English, L.~N.
  Pfeiffer, and K.~W. West.
\newblock {\em Phys. Rev. Lett.}, 61:2701, 1988.

\bibitem{Marm92}
I.~K. Marmorkos and S.~{Das Sarma}.
\newblock {\em Phys. Rev. B}, 45:13396, 1992.

\bibitem{Kall84}
C.~Kallin and B.~I. Halperin.
\newblock {\em Phys. Rev. B}, 30:5655, 1984.

\bibitem{girvin85}
S.~M. Girvin, A.~H. MacDonald, and P.~M. Platzman.
\newblock {\em Phys. Rev. Lett.}, 54:581, 1985.

\bibitem{girvin86}
S.~M. Girvin, A.~H. MacDonald, and P.~M. Platzman.
\newblock {\em Phys. Rev. B}, 33:2481, 1986.

\bibitem{davies97}
H.D.M. Davies, J.C. Harris, J.F. Ryan, and A.~J. Turberfield.
\newblock {\em Phys. Rev. Lett.}, 78:4095, 1997.

\bibitem{Kang01}
{Moonsoo Kang}, Aron Pinczuk, B.~S. Dennis, L.~N. Pfeiffer, and K.~W. West.
\newblock {\em Phys. Rev. Lett.}, 86:2637, 2001.

\bibitem{Luin03}
S.~Luin, V.~Pellegrini, A.~Pinczuk, B.S. Dennis, L.N. Pfeiffer, and K.W. West.
\newblock {\em Phys. Rev. Lett.}, 90:236802, 2003.

\bibitem{jain}
Sudhansu~S. Mandal, Michael~R. Peterson, and Jainendra~K. Jain.
\newblock {\em Phys. Rev. Lett.}, 90:106403, 2003.

\bibitem{Boebinger}
G.~S. Boebinger, H.~W. Jiang, L.~N. Pfeiffer, and K.~W. West.
\newblock {\em Phys. Rev. Lett.}, 64:1793, 1990.

\bibitem{JPE}
M.~Kellogg, J.P.~Eisenstein I.B.~Spielman, L.~N. Pfeiffer, and K.~W. West.
\newblock {\em Phys. Rev. Lett.}, 88:126804, 2002.

\bibitem{Butov}
L.V. Butov.
\newblock {\em Solid State Communication}, 2003.
\newblock In press.

\bibitem{Snoke}
D.~Snoke, Y.~Liu, S.~Denev, L.N. Pfeiffer, and K.W. West.
\newblock {\em Solid State Communication}, 2003.
\newblock In press.

\bibitem{Fong}
H.F. Fong, B.~Keimer, P.W. Anderson, D.~Reznik, F.~Dogan, and I.A. Aksay.
\newblock {\em Phys. Rev. Lett.}, 75:316, 1995.

\end{thebibliography}

\end{document}